\documentclass[aps,amsmath,amssymb,amsfonts, superscriptaddress,twocolumn]{revtex4}
\usepackage[german,american,english]{babel}
\usepackage{graphicx}
\usepackage{graphics}
\usepackage{dcolumn}
\usepackage{multirow}
\usepackage{bm}
\usepackage{latexsym}
\usepackage{amssymb}
\usepackage{amsmath}
\usepackage{amsfonts}
\usepackage{layout}
\usepackage{verbatim}
\usepackage{epsfig}
\usepackage{graphicx}
\usepackage{amsbsy}
\usepackage{xcolor}

\usepackage[caption=false]{subfig}
\newcommand{\bea}{\begin{eqnarray*}}
	\newcommand{\eea}{\end{eqnarray*}}
\newcommand{\bne}{\begin{equation*}}
\newcommand{\ede}{\end{equation*}}

\newcommand{\bnen}{\begin{equation}}
\newcommand{\eden}{\end{equation}}
\newcommand{\bean}{\begin{eqnarray}}
\newcommand{\eean}{\end{eqnarray}}
\newcommand{\bsen}{\begin{subequations}}
	\newcommand{\esen}{\end{subequations}}

\newcommand{\bna}{\begin{array}}
	\newcommand{\eda}{\end{array}}
\newcommand{\bnm}{\begin{enumerate}}
	\newcommand{\edm}{\end{enumerate}}

\begin{document}
	
\title{Disorder in the non-linear anomalous Hall effect of $\mathcal{P}\mathcal{T}$-symmetric Dirac fermions}
\author{Rhonald Burgos Atencia}
\affiliation{School of Physics, The University of New South Wales, Sydney 2052, Australia}
\affiliation{ARC Centre of Excellence in Low-Energy Electronics Technologies, UNSW Node, The University of New South Wales, Sydney 2052, Australia}
\author{Di Xiao}
\affiliation{Department of Materials Science and Engineering, University of Washington, Seattle, WA 98195, USA}
\affiliation{Department of Physics, University of Washington, Seattle, WA 98195, USA}
\author{Dimitrie Culcer}
\affiliation{School of Physics, The University of New South Wales, Sydney 2052, Australia}
\affiliation{ARC Centre of Excellence in Low-Energy Electronics Technologies, UNSW Node, The University of New South Wales, Sydney 2052, Australia}

\begin{abstract}
The study of the non-linear anomalous Hall effect (NLAHE) in $\mathcal{P}\mathcal{T}$-symmetric systems has focussed on intrinsic mechanisms. Here we show that disorder contributes substantially to NLAHE and often overwhelms intrinsic terms. We identify terms to zeroth order in the disorder strength involving the Berry curvature dipole, skew scattering and side-jump, all exhibiting a strong peak as a function of the Fermi energy, a signature of interband coherence. Our results suggest NLAHE at experimentally relevant
transport densities in $\mathcal{P}\mathcal{T}$-symmetric systems is likely to be extrinsic.
\end{abstract}

\date{\today}
\maketitle

\textit{Introduction}. The past decade has witnessed the prediction and observation of the nonlinear anomalous Hall effect (NLAHE), an anomalous Hall response second order in the applied electric field.  Initially motivated by the identification of a Hall response in the absence of time-reversal symmetry breaking~\cite{SodemannPRL2015}, the bulk of research to date has focused on non-magnetic materials \cite{HabibRostami2020,Nagaosa2020,ZhouBenjamin2020,Esin2021,ZiShanLiao2021,ShuaiLi2021,YuanDongWang2022,DaKunZhou2022,NKheirabadi2022,YangWang2022,HuJinXin2022,ZhangChengPing2022,RestaRaffaele2022,Chakraborty_2022,Kaplan2023NatureComm,GholizadehSina2023,ZhengYangZhuang2023,XingGuoYe2023,YueXinHuang2023, MaQiong2019, KaifeiKang2019, ZhihaiHe2021, ZhangCheng2022npj, PanHe2022, SubhajitSinha2022, DuanJunxi2022, LujinMin2023}. 
However, in recent years there has been growing interest in the NLAHE in time-reversal breaking systems such as antiferromagnetic metals~ \cite{Jungwirth2016, Jungwirth2018,Godinho2018,CongXiao2019,ChongWangPRL2021,MazzolaFederico2023}. In particular, in systems with broken time-reversal ($\mathcal T$) and inversion ($\mathcal P$) symmetry, but with the combined $\mathcal{PT}$ symmetry, the linear anomalous Hall effect vanishes and the NLAHE provides the leading contribution to the Hall response.  Thus, while eliciting strong interest for applications in antiferromagnetic spintronics, the NLAHE also provides a tool for the investigation and classification of states with broken symmetries \cite{Jungwirth2016, GaoYang2018, Watanabe2020, Shao2020, ChongWangPRL2021, HuiyingLiu2021}.


In a $\mathcal{PT}$-symmetric system, the leading order contribution to the NLAHE is of order $\tau^0$, where $\tau$ is the relaxation time.  This is in sharp contrast to the $\mathcal T$-symmetric systems where the leading order contribution begins at order $\tau$.  So far the intrinsic contribution has been considered as the only mechanism active at order $\tau^0$~\cite{CongXiao2019, ChongWangPRL2021, HuiyingLiu2021}.  On the other hand, it is well known from the study of the linear anomalous Hall effect that extrinsic contributions such as skew scattering and side jump also manifest at order $\tau^0$, which can compensate or even cancel the intrinsic contribution. Some of these have been addressed in $\mathcal{P}\mathcal{T}$-broken systems \cite{DuZZ2018, Zhang_2018,DuZZ2019,OMatsyshyn2019, SNandy2019,DuZZ2021,OrtixCarmine2021, DuZZ2021NatureComm,SamalSai2021,KaplanDaniel2023}. Yet, to our knowledge, the role of disorder in the NLAHE in $\mathcal{PT}$-symmetric systems has thus far been neglected. This omission is difficult to justify: when seeking to extract intrinsic topological quantities from experimental data the effect of disorder must be incorporated. 

\begin{figure}[hb]
\centering
\includegraphics[width=0.43\textwidth]{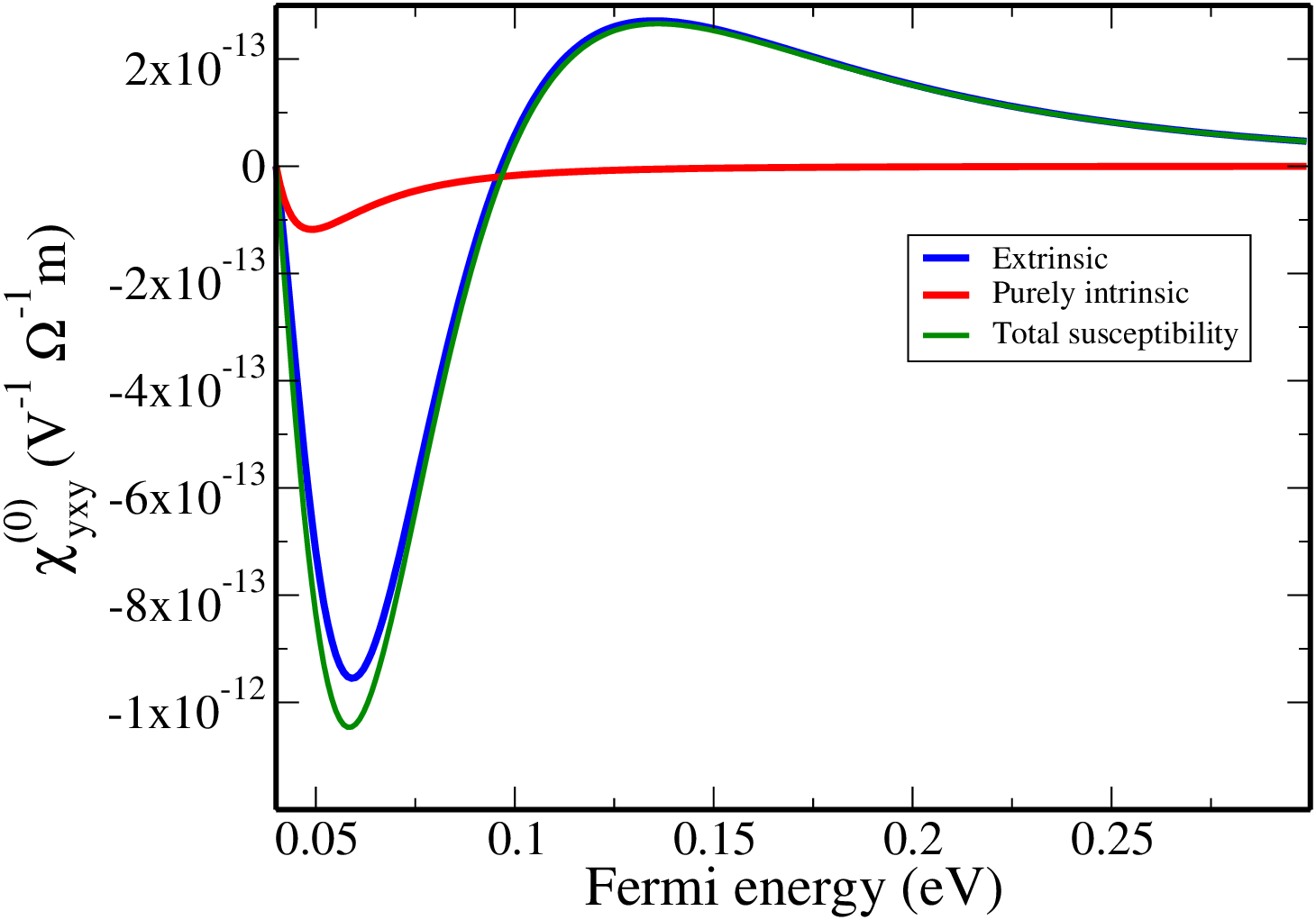}
\caption{
Susceptibility $\chi^{(0)}_{yxy} \propto \tau^0$ with gap $\Delta=40$meV, tilt $t=0.4$ and $v_F=1.6\times 10^{6}$m/s. We approximate the Fermi velocity to be the same for all components, $v_{0x}=v_{0y}$.}
\label{Fig:totalzerotau}
\end{figure}

In this paper we determine the full expression for the NLAHE in the presence of disorder in systems with combined $\mathcal{P}\mathcal{T}$ symmetry, taking 2D tilted Dirac fermions as a prototype system. Defining the non-linear current density $j_i = \chi_{ijk}E_jE_k$, with $\chi_{ijk}$ the non-linear susceptibility, joint $\mathcal{P}\mathcal{T}$-symmetry restricts the powers of $\tau$ that may appear in $\chi$. In a $\mathcal{P}\mathcal{T}$-symmetric system the allowed response scales with even powers of $\tau$, and the susceptibility may be written as $[\chi_{ijk}]=[\chi^{(-2)}_{ijk}] + [\chi^{(0)}_{ijk}]$, where the superscripts indicate the second order and zeroth order in $\tau$ respectively. Because current NLAHE experiments use moderately conducting channels the $\tau^0$ contribution is the most important, and our effort focuses primarily on $\chi^{(0)}_{ijk}$, where disorder competes directly with the intrinsic band structure contributions \cite{ChongWangPRL2021, HuiyingLiu2021,GaoAnyuan2023}. To second order in the electric field disorder contributions involve a complex interplay between band structure and disorder mechanisms. Our central result may be summarized in Fig.~\eqref{Fig:totalzerotau} as the quantitative comparison between intrinsic and disorder contributions.  
Our main findings are as follows. (i) For realistic parameters \cite{CongXiao2019,HuiyingLiu2021} the disorder contributions generally overwhelm the intrinsic terms. This is evident from Fig. \ref{Fig:totalzerotau}, where the total susceptibility essentially tracks the extrinsic contribution. (ii) The NLAHE exhibits a strong peak as a function of the Fermi energy $\varepsilon_F$, whose location is determined by the size of the gap. This peak is present in both $[\chi^{(0)}_{ijk}]$ and $[\chi^{(-2)}_{ijk}]$, and appears in all contributions to the NLAHE, whether intrinsic or induced by disorder. It is a signature of interband coherence \cite{Culcer2017}, a factor that unifies all NLAHE mechanisms. (iii) We identify three main disorder contributions: skew scattering, side jump, and a contribution we term the \textit{extrinsic Berry curvature dipole}, which has not been found previously. It consists of the Berry curvature dipole multiplied by a disorder term that is formally of zeroth order in $\tau$. All the disorder terms are a consequence of an electric field correction to the collision integral. For our prototype model of 2D tilted Dirac fermions all non-linear mechanisms are traced to the Fermi surface. Our quantum mechanical formalism also reveals the existence of additional intrinsic terms that a naive application of the semi-classical method misses, in analogy with Ref.~\cite {lahiri2023intrinsic}.

\begin{figure}[tbp]
\centering
\includegraphics[width=0.43\textwidth]{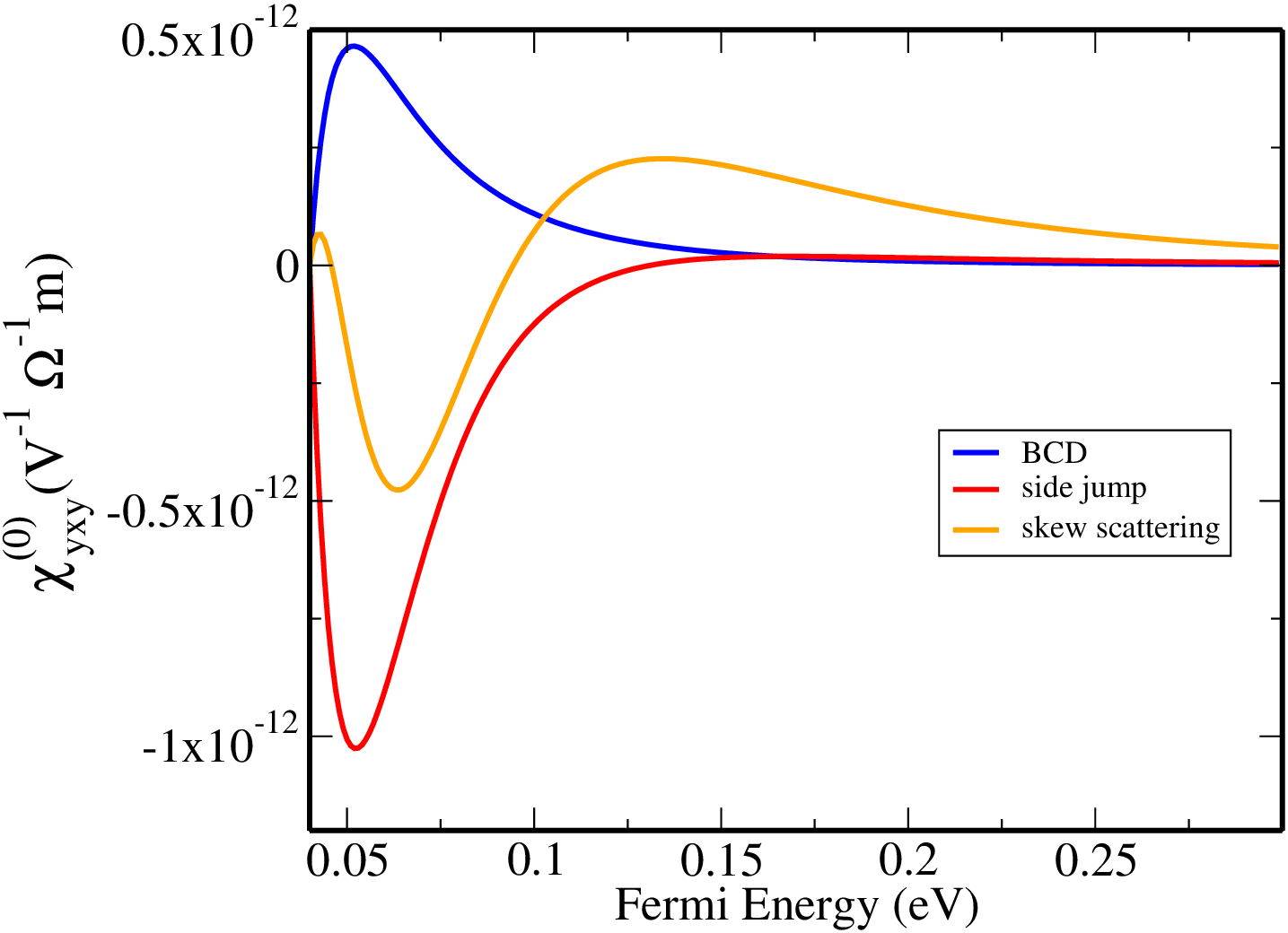}
\caption{Susceptibility $[\chi^{(0)}_{yxy}]$ with gap parameter $\Delta=40$meV and tilt $t=0.4$. $v_F=1.6\times 10^{6}$m/s. We show $[\chi^{(0)}_{yxy}]_{BCD}$, $[\chi^{(0)}_{yxy}]_{sj}$ and $[\chi^{(0)}_{yxy}]_{sk}$.}
\label{Fig:taoindependentDisorder}
\end{figure}

Our findings suggest that the NLAHE signal at experimentally relevant transport densities is dominated by disorder, making an understanding of disorder indispensable in interpreting experimental data. They also provide a strong contrast with $\mathcal{P}\mathcal{T}$-breaking systems studied so far. In that case, with $\mathcal{T}$ preserved, $\mathcal{P}\mathcal{T}$ is necessarily broken in the second-order electrical response, and $\chi \propto \tau$. The NLAHE driven by the Berry curvature dipole (BCD) belongs to this category \cite{SodemannPRL2015,Shao2020,SnehasishNandy2021}, and it was shown that disorder makes a contribution similar in magnitude to the intrinsic terms, without overwhelming them \cite{SnehasishNandy2021}.

\textit{Quantum kinetic equation}. The system is described by the density matrix $\rho(t)$, which obeys the quantum Liouville equation $\partial \rho/\partial {t} + (i/\hbar) \, [H,\rho] = 0$. The Hamiltonian has the form $H = H_0 + e{\bm E}\cdot{\bm r} + U(\bm r)$, with $H_0$ the band Hamiltonian, ${\bm E}$ a constant, uniform electric field, and $U(\bm r)$ the disorder scattering potential. We work in the crystal momentum representation spanned by Bloch states $|m,\bm k\rangle=e^{i\bm k \cdot \bm r}|u^{m}_{\bm k}\rangle$. The disorder model is defined through its correlations functions $\langle U(\bm r)\rangle =0$ and $\langle U(\bm r)U(\bm r')\rangle =u^2_{0}\delta(\bm r-\bm r')$, where $u^2_{0}$ quantifies the strength of disorder. Alternatively, the disorder strength can be measured by 
the momentum relaxation time $1/\tau=\pi \rho(\epsilon_{\bm k})u^2_{0}/\hbar$, with $\rho(\epsilon_{\bm k})$ the density of states.

\textit{Quantum kinetic equation}. Following the methodology of Refs.~\cite{Culcer2017,Sekine2017,BurgosPRR2022}, the density matrix is decomposed into a disorder-averaged part $f_{\bm k}$ and be the focus of our attention, and a fluctuating part, which is integrated out to yield the scattering term in the Born approximation, assuming its time evolution to be Markovian. In this way we obtain the quantum kinetic equation
\begin{equation}
\label{QKE:MainEquationGeneral}
\frac{\partial f}{\partial t} + \frac{i}{\hbar} \, [H_{0}, f] + J_{0}(f) = 
\frac{e{\bm E}}{\hbar}\cdot\frac{Df}{D{\bm k}} 
- J_E(f) - J_{E2}(f).
\end{equation}
The covariant derivative appearing above reads $\frac{Df_{\bm k}}{D{\bm k}} = \frac{\partial f_{\bm k}}{\partial{\bm k}} 
-i[\bm{\mathcal{R}}_{\bm k}, f_{\bm k}]$, with the Berry connection $\bm{\mathcal{R}}^{mm'}_{\bm k}=i\langle u^{m}_{\bm k}| \nabla_{\bm k} u^{m'}_{\bm k}\rangle$. The covariant derivative accounts for the momentum dependence of the basis functions. The density matrix $f_{\bm k}$ has both diagonal and off-diagonal elements in band index $m$. We represent the band-diagonal part by $n_{\bm k}$ and the off-diagonal part by $S_{\bm k}$, such that $f_{\bm k} = n_{\bm k} + S_{\bm k}$. The equilibrium density matrix is band-diagonal with matrix elements given by the Fermi-Dirac distribution $n_{FD}(\epsilon^{m}_{\bm k})$ for each band.

\begin{figure}[tbp]
\centering
\includegraphics[width=0.43\textwidth]{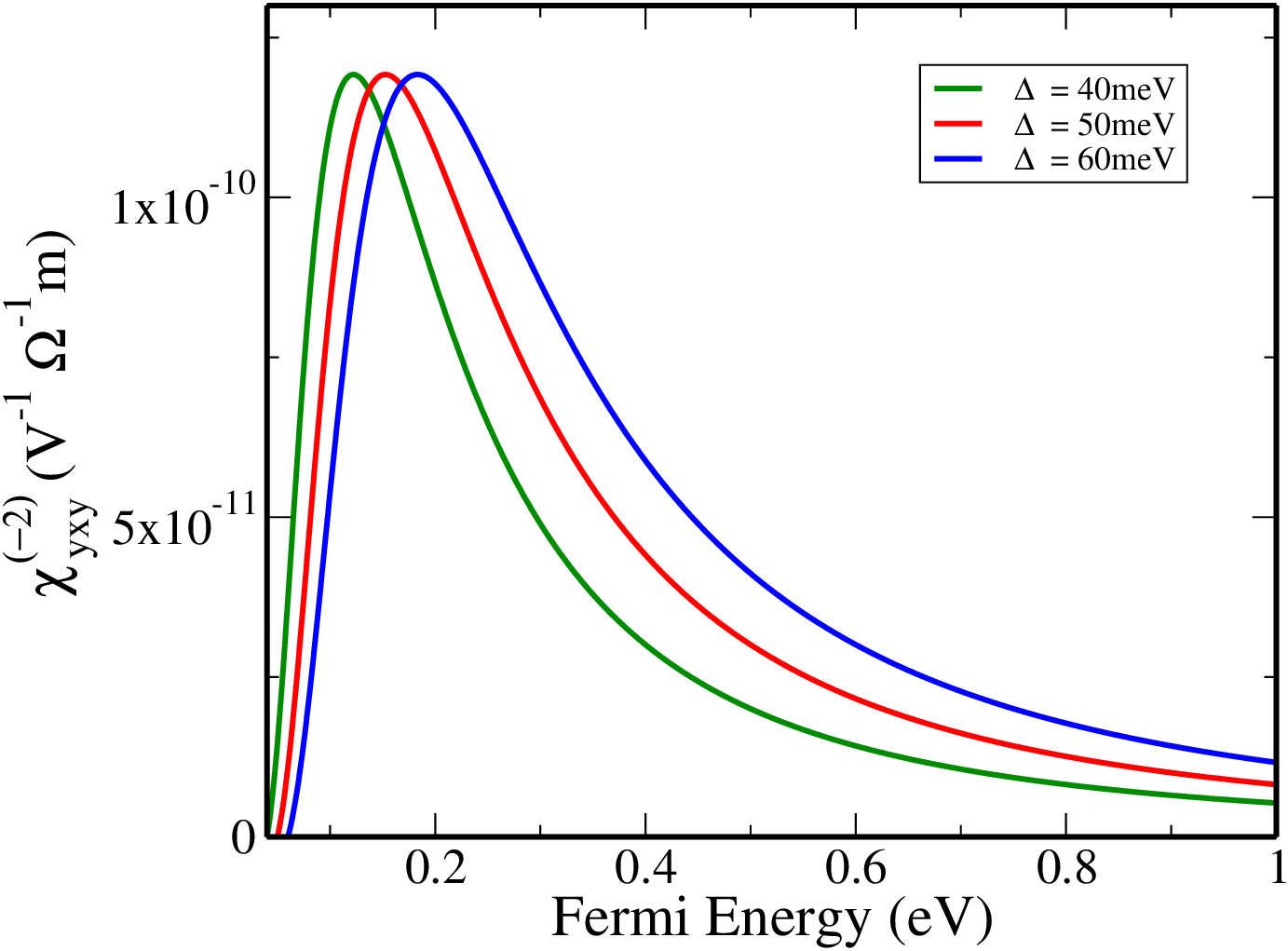}
\caption{Leading order susceptibility $\chi^{(-2)}_{yxy} \propto \tau^2$ with $t=0.4$, $v_F=1.6\times 10^{6}$m/s and $\tau=1$ps. We have approximated the Fermi velocity to be the same for all components.}
\label{Fig:Leanding}
\end{figure}

The bare collision integral is defined as $J_{0}(f) = (i /\hbar) \langle [U,g_{0}] \rangle$ with the function 
\begin{equation}
g_0=\frac{1}{2\pi i}\int^{\infty}_{-\infty}d\epsilon G^{R}_{0}(\epsilon)[ U  , f ] G^{A}_{0}(\epsilon) 
\end{equation}
and the the electric field correction $J_{E}(f)$ defined as 
\begin{equation}
\label{}
J_E(f)= \frac{1}{2\pi\hbar }\int^{\infty}_{-\infty}d\epsilon \langle [U, G^{R}_{0}(\epsilon)[e{\bm E}\cdot{\bm r}, g_0 ] G^{A}_{0}(\epsilon)] \rangle. 
\end{equation}
The retarded Green's function is defined as $G^{R}_0(\epsilon)=-\frac{i}{\hbar}\int^{\infty}_{0} dt e^{-iH_0t/\hbar} e^{i\epsilon t/\hbar} e^{-\eta t}$, where the factor $e^{-\eta t}$ ensures convergence and the advanced Green's function $G^{A}_0(\epsilon)$ follows by Hermitian conjugation.
The collision integral $J_{E2}(f) \rightarrow J_{E2}(n_{FD})$ is by itself second order in electric field and must be evaluated as a functional of the equilibrium Fermi-Dirac distribution. Its main role is to eliminate Fermi sea effects in the same way as in linear response -- this is explained in the Supplement. 

We solve Eq.\eqref{QKE:MainEquationGeneral} perturbatively in the electric field and disorder strength quantified by $\tau$. The leading order correction in linear response comes from the first driving term on the RHS by taking $f\rightarrow n_{FD}$. It will give the Boltzmann-like contribution $n^{(-1)}_{E\bm k}$, which in our notation refers to a band-diagonal term $\propto \tau$. The kinetic equation is solved iteratively in ${\bm E}$ up to second order, which is denoted by the subscript $E2$. As an example, the leading second order response follows from the same driving term by using the linear Boltzmann equation. It will produce a contribution $n^{(-2)}_{E2\bm k}$, i.e. band-diagonal, second order in ${\bm E}$ and quadratic in $\tau$. We will use a similar notation to represent the off-diagonal channel. Once the distribution is determined, the 
current follows from the trace $j=-e \displaystyle \sum_{\bm k,mm'} [v^{mm'}_{\bm k} f^{m'm}_{E2\bm k}]$, namely, the velocity operator weighted by the density matrix.

\textit{Model Hamiltonian}. We investigate a system that breaks both time reversal $\mathcal{T}$ and parity $\mathcal{P}$ symmetry but preserves the joint $\mathcal{P}\mathcal{T}$ symmetry. A generic paradigm is provided by a tilted Dirac cone. The band Hamiltonian for a single valley has the form 
\begin{equation}
H_0= \hbar v_t k_x\sigma_0 + \hbar v_{0x}k_x \sigma_x \pm \hbar v_{0y}k_y \sigma_y + \Delta \sigma_z,    
\end{equation}
where the first term is the tilt, $\sigma_{i}$ are Pauli matrices, $v_{0i}$ are Fermi velocities and the term $\Delta$ is the energy gap. We replace $k_i \rightarrow v_{0i}k_i$ with the following rule for integrals $\sum_{\bm k} (\cdots)\rightarrow \frac{1}{(2\pi)^2}\frac{1}{v_{0x}v_{0y}}\int dk_x dk_y (\cdots)$ and derivatives $\frac{\partial}{\partial k_i} \rightarrow v_{0i} \frac{\partial}{\partial k_i}$. Since we assume the two $\mathcal{P}\mathcal{T}$-symmetric states can be decoupled, we focus on the Hamiltonian with the positive sign. The eigenvalues read $\epsilon^{\pm}_{\bm k}
=\hbar t k_x \pm  \epsilon_{0k}$ with $\epsilon_{0k} =\sqrt{\hbar^2k^2_x + \hbar^2k^2_y+\Delta^2}$ and the dimensionless parameter $t=v_t/v_{0x}$. This dimensionless parameter controls the breaking of inversion symmetry necessary for the nonlinear response, hence the non-linear susceptibility will be at least of first order in $t$.

\textit{Results}. The components of the non-linear susceptibility are related by $[\chi^{(0)}_{xxx}]=[\chi^{(0)}_{xyy}]+[\chi^{(0)}_{yxy}]$, apart from a proper velocity prefactor, meaning $[\chi^{(0)}_{xxx}] \propto v^3_{0x}$, while the right hand side is $ \propto v_{0x}v^2_{0y}$. The first index in $\chi$ is the direction of the current while the last two indices represent the two factors of the electric field. Below we show results for $ [\chi^{(0)}_{yxy}] = [\chi^{(0)}_{xxx}]-[\chi^{(0)}_{xyy}] $. 



Solving the kinetic equation to zeroth order in $\tau$ we identify a purely intrinsic contribution as well as three disorder corrections to the non-linear anomalous Hall response of $\mathcal{P}\mathcal{T}$-symmetric systems: a side jump effect, a skew scattering effect and a Berry curvature dipole effect. The Berry curvature dipole and side jump contributions emerge in the off-diagonal channel of the density matrix and follow from the equation 
\begin{align}
\label{QKE-2nd-ODiagonal-0}
&\frac{\partial S^{(0)mm'}_{E2{\bm k}}}{\partial t} + \frac{i}{\hbar} \, [H_{0{\bm k}}, S^{(0)}_{E2{\bm k}}]^{mm'} 
= 
\frac{e{\bm E}}{\hbar}\cdot\frac{D S^{(0)mm'}_{E \bm k,int }}{D \bm k} \nonumber \\
&-i\frac{e{\bm E}}{\hbar}\cdot [ \bm{\mathcal{R}}_{\bm k}, n^{(0)}_{E\bm k,sj}]^{mm'}  - [J_{0}(n^{(-1)}_{E2,sj})]^{mm'}_{\bm k}.
\end{align}
The general solution reads $S^{(0)mm'}_{E2{\bm k}}=-i \hbar (\epsilon^{m}_{\bm k}-\epsilon^{m'}_{\bm k})^{-1}  d^{(0)mm'}_{E2{\bm k}}$, where $ d^{(0)mm'}_{E2{\bm k}}$ refers generically to the driving term on the right hand side of the equation.

The first term in Eq.\eqref{QKE-2nd-ODiagonal-0} produces the second order intrinsic distribution. It is related to the covariant derivative of the intrinsic linear response 
that follows from the equation $S^{(0)mm'}_{E \bm k,int }=-(\epsilon^{m}_{\bm k}-\epsilon^{m'}_{\bm k})^{-1}
e{\bm E} \cdot [ \bm{\mathcal{R}}_{\bm k} , n_{FD} ]^{mm'} $. This is the only intrinsic contribution to the non-linear response in the sense that it depends solely on the band structure. After tracing the off-diagonal velocity $v^{mm'}_{\bm k, i}=i\hbar^{-1} (\epsilon^{m}_{\bm k} - \epsilon^{m'}_{\bm k'} ) \mathcal{R}^{mm'}_{\bm k, i} $ 
with the intrinsic distribution we obtain the susceptibility 
\begin{align}
\label{Eq:Purely-Intrisic-Contributio}
[\chi^{(0)}_{yxy}]_{int}
&=-
\frac{t}{8} \hbar^2 e^3 v^2_{0y}  v_{0x} 
\frac{\rho(\epsilon_{F}) }{\epsilon^3_{F}} \xi^2_{F} 
( 1-\xi^2_{F}),
\end{align}
where $\rho(\epsilon_{F})$ is the density of states and we defined the parameter $\xi_{F}=\Delta/\epsilon_{F}$. This contribution is an inter-band coherence effect where virtual transitions between valence and conduction band are mediated by the product of off-diagonal terms in the Berry connection. It is also a Fermi surface response, vanishing at $\xi_{F}=1$. This is in contrast to the intrinsic linear anomalous Hall effect, which is a Fermi sea response. 

We turn our attention to the disorder corrections to the susceptibility. The band-diagonal term to zeroth order in $\tau$ reads $n^{(0)++}_{Ey\bm k,sj}=- e E_{y}v_{0y}A_{0}(\bm k) + \cdots $, where the coefficient is 
$A_0(\bm k) = t\frac{\tau_{sp}\hbar }{2 \tau \epsilon_{0\bm k}}
\xi_k\left[(1+\xi^2_{k}) 
+ (1-\xi^2_{k})\epsilon_{0\bm k}\frac{\partial  }{\partial \epsilon^{+}_{0\bm k} }  \right] \delta(\epsilon^{+}_{0\bm k}-\epsilon_{F}) $. We have ignored higher harmonics irrelevant for transport. We have defined the transport time and the single particle relaxation time as 
$1/\tau_{tr}=(1+3\xi^2_{k})/2\tau$ and $1/\tau_{sp}=(1+\xi^2_{k})/\tau$ respectively.
Tracing the off-diagonal velocity with this channel of the off-diagonal density matrix will produce 
a Berry curvature dipole (BCD) like contribution given by the current $j_{i}=- (e^2/\hbar)\sum_{\bm k,m}\left( \bm E \times {\bm\Omega}^{mm}_{\bm k} \right)_{i} n^{(0)mm}_{E\bm k}$ where 
${\bm\Omega}^{mm}_{\bm k}$ is the Berry curvature. Explicit evaluation yields
\begin{align}
\label{ }
[\chi^{(0)}_{yxy}]_{BCD}
&= \frac{t e^3 \hbar^2 v_{0x} v^2_{0y}}{2}  
\frac{\rho(\epsilon_{F})\xi^2_{F}}{\epsilon^3_{F}}
\frac{(1-\xi^2_{F}) (2+\xi^2_{F}) }{(1+\xi^2_{F})^2}.
\end{align}
This contribution is shown in Fig. \eqref{Fig:taoindependentDisorder} and given in full in the Supplement. The BCD susceptibility $[\chi^{(0)}_{yxy}]_{BCD}$ is the analogue of the 
BCD widely studied in $\mathcal{P}\mathcal{T}$-broken systems. However, the BCD appears here as an inter band coherence effect involving \textit{disorder}, and we refer to this contribution as the \textit{extrinsic Berry curvature dipole}. It is a consequence of the driving term arising from the electric field corrected collision integral. This response is also a Fermi surface effect.

Let us consider the last driving term in the kinetic equation. The first order in $\tau$ distribution in the collision integral in eq. \eqref{Eq:Purely-Intrisic-Contributio} reads
\begin{widetext}
\begin{align}
\label{Eq:SecondSideJumpEffectLinearTau}
n^{(-1)++}_{xy\bm k,sj}&=-\frac{e^2}{\hbar}v_{0y}E_{y}v_{0x}E_{x}
\tau_{tr} \cos(\theta_{\bm k})  \frac{\partial A_{0}(\bm k) }{\partial k}  + ...  
\end{align}
Tracing the off-diagonal velocity with $S^{(0)mm'}_{E2{\bm k}}=i \hbar (\epsilon^{m}_{\bm k}-\epsilon^{m'}_{\bm k})^{-1}  [J_{0}(n^{(-1)}_{E2,sj})]^{mm'}_{\bm k}$ yields
\begin{align}
\label{Eq:SecondSideJumZeroOrder}
[\chi^{(0)}_{yxy}]^{(od)}_{sj}
&=
\frac{1}{2}te^3 v^2_{0y}v_{0x}\hbar^2
 \frac{  \rho(\epsilon_{F}) }{\epsilon^3_{F}} 
 \xi^2_{F}(1-\xi^2_{F})
 \Lambda_s \Lambda_t
 \left\{ (1-\xi^2_{F})^2 \Lambda_s
 -
 8\xi^2_{F}(1-\xi^2_{F}) \Lambda_s \Lambda_t 
 \nonumber \right. \\
 &\left. 
 +
 3(1-\xi^2_{F})  
 -24\xi^2_{F} \Lambda_t
 - 24\xi^2_{F}(1-\xi^2_{F}) \Lambda_t^2
 \right\}.
 \end{align} 
\end{widetext}
The dimensionless parameters $\Lambda_s=(\tau_{sp}/\tau)$ and $\Lambda_t=(\tau_{tr}/2\tau)$. This coefficient represent the side jump in analogy to linear response \cite{BurgosPRR2022}. 
It is a Fermi surface response, as well as an inter-band coherence effect due to virtual transitions mediated by the Berry connection. 

In the band-diagonal part of the density matrix to zeroth order $\tau$ we identify a skew scattering contribution and a second side jump contribution. They follow from
$[J(n^{(0)}_{E2})]^{mm}_{\bm k} = -[J_{0}(S^{(0)}_{E2}) ]^{mm}_{\bm k}- [ J_{E}(n^{(0)}_{E,sj}) ]^{mm}_{\bm k} $, where the driving terms are to the right. 
Skew scattering follows from the first driving term  
while the second gives the side-jump contribution. This is identical to Eq. \eqref{Eq:SecondSideJumZeroOrder}, hence the electric field corrected collision integral doubles the side jump, in analogy with linear response \cite{BurgosPRR2022, Culcer2010}. 
We plot this side jump term in Fig. \eqref{Fig:taoindependentDisorder}.

Finally, the skew scattering contribution is found as 
\begin{widetext}
\begin{align}
\label{}
[\chi^{(0)}_{yxy}]_{sk}
&=
e^3 v^2_{0y}v_{0x} 
\frac{3}{2} t \hbar^2
\frac{ \rho(\epsilon_{F})}{\epsilon^3_{F}} 
\xi^2_{F}(1-\xi^2_{F})
\Lambda_t^2
\left\{ 
3(1-\xi^2_{F}) 
-16 \xi^2_{F} \Lambda_t
+15(1-\xi^2_{F})^{2} \Lambda_t 
-112\xi^2_{F}(1-\xi^2_{F}) \Lambda_t^2
\right. \nonumber \\
&\left. 
-12 \xi^2_{F}(1-\xi^2_{F}) \Lambda_s \Lambda_t
+64\xi^4_{F}(1+\xi^2_{F}) \Lambda_s \Lambda_t^2 
-48 \xi^4_{F} (1-\xi^2_{F})^2 \Lambda_t^3 
+128\xi^6_{F} \Lambda_t^3
\right\}.
\end{align}
\end{widetext}
Although this quantity exhibits a similarly complex dependence on the Fermi energy, one can straightforwardly identify it as a Fermi surface effect. It represents inter-band coherence mediated by disorder through an extrinsic off-diagonal term in the density matrix. It is shown in Fig. \eqref{Fig:taoindependentDisorder}. 
All contributions exhibit similar behaviour, namely, $\propto 1/\epsilon^{4}_{F}$ that dominates for increasing Fermi energy and $\propto (1-\Delta^2/\epsilon^{2}_{F})$ that
makes all the expressions zero when the Fermi energy approaches the gap (a signature of Fermi surface effect in the prefactor of all the contributions we have calculated).
This non-monotonic behaviour of the susceptibility 
makes the function to bent in between these two limiting cases and to develop a peak. The sign of the peak is dictated by the dominant term in the 
terms inside the curly braces in each expression. For instances, for the side jump susceptibility, the second and last terms scale similar to the pre-factor 
and make the function to develop a negative peak. It is similar for the skew scattering contribution. The peak develops in the vicinity of the gap, as the bands approach each other, revealing the effect of interband coherence. In fact, since this behavior is shared by all intrinsic and extrinsic contributions, we regard interband coherence as the unifying physical mechanism behind the NLAHE.



To fully account the transversal susceptibility of $\mathcal{P}\mathcal{T}$-symmetric Dirac fermions, we also solved the Boltzmann like equation   
$[J_{0}(n^{(-2)}_{E2})]^{mm}_{\bm k} = \frac{e{\bm E}}{\hbar}\cdot\nabla_{\bm k}n^{(-1)mm}_{E{\bm k}}$ for the leading order susceptibility. Its behaviour is shown in Fig. \eqref{Fig:Leanding}. It is also Fermi surface effect that vanishes when we approach the gap and also vanishes in time reversal symmetric systems. The peak shifts as a function of the gap, with a similar behavior noted in $[\chi^{(0)}_{ijk}]$.

\textit{Conclusions}. We have calculated the electrical susceptibility to second order in the electric field in $\mathcal{P}\mathcal{T}$-symmetric 2D tilted Dirac fermions. We have demonstrated the existence of intra- and inter-band disorder effects that are counterparts of the side jump and skew scattering terms in linear response, as well as a new Berry curvature dipole correction, which is disorder-dependent yet appears at zeroth order in the disorder strength. We showed that disorder corrections generally overwhelm the intrinsic contribution and are expected to play a vital role in realistic samples, where disorder is unavoidable. 


\textit{Acknowledgments}. This work is supported by the Australian Research Council Centre of Excellence in Future Low-Energy Electronics Technologies, project number CE170100039. D.X.~is supported by AFOSR MURI 2D MAGIC (FA9550-19-1-0390).

\bibliography{NLHE}

\end{document}